\documentclass[submission,copyright,creativecommons]{eptcs}
\providecommand{\event}{EXPRESS/SOS 2019}

\usepackage{underscore}
\usepackage{xspace}
\usepackage{amsfonts,amssymb,amsmath}
\usepackage{stmaryrd}
\usepackage{nicefrac}
\usepackage{paralist}

\usepackage{Encodings}

\title{Comparing Process Calculi Using Encodings}
\def\titlerunning{Comparing Process Calculi Using Encodings}

\author{Kirstin Peters\institute{TU Berlin/TU Darmstadt}}
\def\authorrunning{K. Peters}

\begin{document}

\maketitle

\begin{abstract}
	Encodings or the proof of their absence are the main way to compare process calculi. To analyse the quality of encodings and to rule out trivial or meaningless encodings, they are augmented with encodability criteria.
	There exists a bunch of different criteria and different variants of criteria in order to reason in different settings. This leads to incomparable results.
	Moreover, it is not always clear whether the criteria used to obtain a result in a particular setting do indeed fit to this setting.
	This paper provides a short survey on often used encodability criteria, general frameworks that try to provide a unified notion of the quality of an encoding, and methods to analyse and compare encodability criteria.
\end{abstract}


\section{Introduction}

Process calculi (or process algebra) is one area of formalisations of concurrent systems. Other areas are for example Petri nets \cite{petri62} or the Actor model \cite{hewittBishopSteiger73}. \emph{A brief history of process algebra} can be found in \cite{baeten05}.
Over the time different process calculi emerge. Examples are \ccs \cite{milner89}, the \piCal \cite{milnerParrowWalker92, milner99}, \csp \cite{hoare78}, or \acp \cite{bergstraKlop82}, just to name some of the most prominent ones. Note that each of these calculi denotes rather a family of process calculi.
The number of different process calculi is in fact enormous. As discussed in \cite{nestmann06} there are many good reasons for this great number of different calculi. The most plausible is maybe that many of these different calculi stem from different practical needs.
They are designed as domain-specific calculi capturing exactly the set of primitives necessary to model the desired system at a proper level of abstraction without overloading the theory with (for this domain) unnecessary operations.
The large number of calculi calls for methods to compare different calculi or different variants within a family of process calculi with respect to their expressive power. The most prominent such method is language embedding using encodings \cite{boerPalamidessi1991}.
Encodings---or the proof of their absence---do not only allow to compare the expressive power of languages but also formalise similarities and differences between the considered languages. So they provide a base for implementations of languages into real systems.

There are basically two ways to measure the expressive power of a language \cite{parrow08}.

An \emph{absolute result} is a result about the expressive power of a single process calculus, usually obtained by proving the ability (\emph{positive absolute result}) or inability (\emph{negative absolute result}) to solve some kind of problem (see \cite{parrow08, gorla10} and even \cite{liptonSnyderZalcstein74}), whereas a \emph{relative result} compares two process calculi.

To \emph{compare} the absolute expressive power of \emph{two} languages, we may simply choose a problem that can be solved in one language, but not in the other language. Note that combining two absolute results that are both positive or both negative usually does not reveal much information, because it proves only that the considered languages do not differ with respect to the respective particular problem. Actually as soon as we compare two languages, it makes sense to use the term \emph{relative expressive power}, as we can now relate the two languages. The terminology of the relative expressive power has also been attributed (see~\cite{parrow08, gorla10}) to the comparison of the expressive power of two languages by means of the existence or non-existence of encodings from one language into the other language, subject to various conditions on the encoding. Indeed, encodings are the main way to relate the expressive power of two process calculi. A positive relative result, \ie the proof of the existence of an encoding, is denoted as \emph{encodability result}, whereas a negative result is denoted as \emph{separation result}.

Language comparison by means of encodings is a wide area of research within the context of process calculi. Reasonable and meaningful encodings from one language into another show that the latter is at least as expressive as the former, whereas the absence of such an encoding shows that the former can express some behaviour that is not expressible in the latter, \ie reveals a difference in the expressive power of the former compared to the latter language. To analyse the quality of encodings and to rule out trivial or meaningless encodings, they are augmented with encodability criteria. However, as stated several times in literature (\eg in \cite{palamidessi03, nestmann06, parrow08, gorla10}), there is no agreement on what set of criteria makes an encoding reasonable and meaningful.
Sometimes it is even stated that such an agreement may not exist or may not be desirable (see \eg \cite{palamidessi03}), because many criteria result from different practical needs. They are often derived from the main purpose of the current analysis. From a practical point of view this is meaningful. But, obviously, using different quality criteria for different results, because they were motivated by different practical problems, naturally leads to incomparable results.

After some technical preliminaries in Section~\ref{sec:techPrel}, Section~\ref{sec:encodabilityCriteria} provides a short survey of often used encodability criteria. This section is the heart of this paper. Then, Section~\ref{sec:frameworks} shortly outlines three approaches that try to overcome the problem of incomparable results and provide a unified notion of the quality of an encoding. Similarly, Section~\ref{sec:evaluation} shortly outlines the only way to formally analyse and compare encodability criteria we are aware of. Finally, Section~\ref{sec:conclusions} raises some open questions.
Note that this paper provides a rather abstract view on encodings and focuses on encodability criteria. A more practical study of encodings, including a discussion of how to obtain an encoding and prove it correct as well as some actual examples of encodings can be found in \cite{peters12}.


\section{Process Calculi and Encodings}
\label{sec:techPrel}

A \emph{process calculus} is a language $ \lang_C = \Tuple{\proc_C, \step_C} $ consisting of a set of terms $ \proc_C $---its \emph{syntax}---and a relation on terms $ \step_C \; \subseteq \proc_C \times \proc_C $---its \emph{semantics}.
The elements of $ \proc_C $ are called \emph{process terms} or shortly processes or terms. We use upper case letters $ P, Q, S, T, \ldots $ to range over process terms.

The \emph{syntax} is usually defined by a context-free grammar defining operators. An \emph{operator} is a function from sorts and process terms into a process term. Sorts can be used to introduce data values or structural information such as location names or identities. For simplicity, we restrict our attention to a single such sort. Assume a countably-infinite set $ \names $, whose elements are called \emph{names}. Names are the universe of elements of which the processes are constructed within many process calculi such as \eg in the \piCal. Then an operator is a function $ \opn : \names^n \times \proc_C^m \to \proc_C $ from names and process terms into a process term. In this case, we say $ \opn $ has the \emph{arity} $ m $. Sometimes, process calculi also specify operators that, instead of a fixed number of arguments, accept any finite set of names and/or process terms usually indexed by a finite index set $ \indexSet $. In this case, the arity of the operator is not a fixed value but, for a given set of arguments, is determined by the number of process terms among the arguments. An example of such an operator is given by the choice operator in the \piCal. Operators of arity $ 0 $ are denoted as \emph{constants}.

The semantics of a process calculus can be given as a reduction semantics---that describes the interaction within a system of processes---or a labelled semantics---that also describes the interactions of such a system with its environment.
Here we assume that the semantics of the language is provided as a reduction semantics, because in the context of encodings the treatment of reductions is simpler.
A \emph{step} $ P \step_C P' $ is an element $ (P, P') \in \; \step_C $. Let $ P \step_C $ denote the existence of a step from $ P $, \ie $ \exists P' \in \proc_C \logdot P \step_C P' $, let $ P \noStep_C $ denote the absence of a step from $ P $, \ie $ \neg\left( P \step_C \right) $, and let $ \steps_C $ denote the reflexive and transitive closure of $ \step_C $. We write $ P \StepsW{C} $ if $ P $ can do an infinite sequence of steps. A term $ P $ such that $ P \StepsW{C} $ is called \emph{divergent}.

Intuitively, a \emph{context} $ \Context{}{}{\hole_1, \ldots, \hole_n} : \proc_C^n \to \proc_C $ is a process term with $ n $ holes. Formally, it is a function from $ n $ process terms into a process term, \ie given $ n $ terms $ P_1, \ldots, P_n \in \proc_C $, the term $ \Context{}{}{P_1, \ldots, P_n} $ is the result of inserting the $ n $ terms $ P_1, \ldots, P_n $ in that order into the $ n $ holes of $ \context $. We also consider contexts $ \Context{}{}{\hole_1, \ldots, \hole_n, \hole_{n + 1}, \ldots, \hole_{n + m}} : \names^n \times \proc_C^m \to \proc_C $ that have holes for names and terms.

A typical operator is the restriction of scopes of names. A \emph{scope} defines an area in which a particular name is known and can be used. For several reasons, it can be useful to restrict the scope of a name. For instance to forbid interactions between two processes or with an unknown and, hence, potentially untrusted environment. Names whose scope is restricted such that they cannot be used from outside the scope are denoted as \emph{bound names}. The remaining names are called \emph{free names}. Accordingly, we assume three sets---the sets of names $ \Names{P} $ and its subsets of free names $ \FreeNames{P} $ and bound names $ \BoundNames{P} $---for each term $ P $. In the case of bound names, their syntactical representation as lower case letter serves as a place holder for any fresh name, \ie any name that does not occur elsewhere in the term.
To avoid name capture or clashes, \ie to avoid confusion between free and bound names or different bound names, bound names can be mapped to fresh names by \emph{$ \alpha $-conversion}. We write $ P \equiv_{\alpha} Q $ if $ P $ and $ Q $ differ only by $ \alpha $-conversion.
A \emph{substitution} $ \sigma $ is a finite mapping from names to names defined by a set $ \Set{ \Subst{y_1}{x_1}, \ldots, \Subst{y_n}{x_n} } $ of renamings, where the $ x_1, \ldots, x_n $ are pairwise distinct. The application of a substitution to a term $ P\Set{ \Subst{y_1}{x_1}, \ldots, \Subst{y_n}{x_n} } $ is defined as the result of simultaneously replacing all free occurrences of $ x_i $ by $ y_i $ for $ i \in \Set{ 1, \ldots, n } $, possibly applying $ \alpha $-conversion to avoid capture or name clashes. For all names $ \names \setminus \Set{ x_1, \ldots, x_n } $ the substitution behaves as the identity mapping.

To compare process terms, process calculi usually come with different well-studied preorders and equivalences (see \cite{vanglabbeek01, vanglabbeek93} for an overview and a classification of the most frequent equivalences).
A special kind of equivalence with great importance to reason about processes are congruences. A congruence is the closure of an equivalence with respect to contexts, \ie an equivalence $ \rel \subseteq \proc_C \times \proc_C $ is a \emph{congruence} if $ \left( P, Q \right) \in \rel $ implies $ \left( \Context{}{}{P}, \Context{}{}{Q} \right) \in \rel $ for all terms $ P, Q \in \proc_C $ and all contexts $ \Context{}{}{\hole} : \proc_C \to \proc_C $.
Moreover, let $ C $ be a set of $ \proc_C \to \proc_C $-contexts, then $ \rel \subseteq \proc_C \times \proc_C $ is a \emph{congruence with respect to} $ C $ if $ \left( P, Q \right) \in \rel $ implies $ \left( \Context{}{}{P}, \Context{}{}{Q} \right) \in \rel $ for all terms $ P, Q \in \proc_C $ and all contexts $ \context \in C $.
Moreover, process calculi usually come with a special congruence $ \equiv_C \; \subseteq \proc_C \times \proc_C $ called \emph{structural congruence}. Its main purpose is to equate syntactically different process terms that model quasi-identical behaviour.

Of special interest are simulation relations; in particular bisimulations \cite{sangiorgi09}.
$ \rel $ is a bisimulation if any two related processes mutually simulate their respective sequences of steps, such that the derivatives are again
related.

\begin{definition}[Bisimulation]
	\label{def:bisimulation}
	$ \rel $ is a \emph{(weak reduction) bisimulation} if for each $ (P, Q) \in \rel $:
	\begin{compactitem}
		\item $ P \steps_C P' $ implies $ \exists Q' \logdot Q \steps_C Q' \wedge (P', Q') \in \rel $
		\item $ Q \steps_C Q' $ implies $ \exists P' \logdot P \steps_C P' \wedge (P', Q') \in \rel $
	\end{compactitem}
	Two terms are \emph{bisimilar} if there exists a bisimulation that relates them.
\end{definition}

\noindent
The definition of a \emph{strong (reduction) bisimulation} is obtained by replacing all $ \steps_C $ by $ \step_C $ in the above definition, \ie a strong bisimulation requires that a step has to be simulated by a single step.
Coupled similarity (defined below) is strictly weaker than bisimilarity. As pointed out in \cite{parrowSjoedin92}, in contrast to bisimilarity it allows for intermediate states in simulations: states that cannot be identified with states of the simulated term. Each symmetric coupled simulation is a bisimulation.

\begin{definition}[Coupled Simulation]
	\label{def:coupledSimulation}
	A relation $ \rel $ is a \emph{(weak reduction) coupled simulation} if, for each $ (P, Q) \in \rel $ with $ P \steps_C P' $, $ \left( \exists Q' \logdot Q \steps_C Q' \wedge (P', Q') \in \rel \right) $ and $ \left( \exists Q' \logdot Q \steps_C Q' \wedge (Q', P') \in \rel \right) $.\\
	Two terms are \emph{coupled similar} if they are related by a coupled simulation in both directions.
\end{definition}

An \emph{encoding} from $ \langS = \Tuple{\procS, \stepS} $ into $ \langT = \Tuple{\procT, \stepT} $ relates two process calculi. We call $ \langS $ the \emph{source} and $ \langT $ the \emph{target language}. Accordingly, terms of $ \procS $ are \emph{source terms} and of $ \procT $ \emph{target terms}. In the simplest case an encoding from $ \langS $ into $ \langT $ is an \emph{encoding function} $ \encoding : \procS \to \procT $ from source terms into target terms.
Sometimes an encoding is defined by several functions, such as the encoding function and the renaming policy (see below).
Else we identify an encoding with its encoding function.

Sometimes it is necessary to translate a source term name into a sequence of names or reserve some names for the encoding function. To ensure that there are no conflicts between these reserved names and the source term names, \cite{gorla10} equips an encoding with a \emph{renaming policy} $ \phi $, i.e., a substitution from names into sequences of pairwise disjoint names.
To keep distinct names distinct, \cite{gorla10} assumes that the sequences of names that result from applying a renaming policy to distinct names have no common name. Moreover, if the renaming policy translates a single name into a sequence of names then the length of such a sequence has to be the same for all names, such that the encoding can not distinguish between different source term names by the length of the sequences to which they are encoded. Obviously, no name should be translated into an infinite sequence of names.

\begin{definition}[Renaming Policy]
	\label{def:renamingPolicy}
	A substitution $ \phi : \names \to \names^n $ from names into sequences of pairwise disjoint names is a renaming policy, if $ \forall x, y \in \names \logdot x \neq y \text{ implies } \phi(x) \cap \phi(y) = \emptyset $, where $ \phi(z) $ is considered as a set.
\end{definition}

\noindent
Note that the renaming policy allows us to use the names reserved by the encoding like implicit parameters. It is for instance possible that some part of the encoding introduces a free occurrence of a reserved name within the encoding of a subterm which is bound by the surrounding part of the encoding. Examples can be found in \cite{gorla10, peters12}. Accordingly, in \cite{gorla10} an encoding is a pair $ \left( \encoding, \phi(\cdot) \right) $.

An \emph{encodability criterion} is a predicate on encoding functions, used to reason about the quality of encodings. To simplify the presentation we assume henceforth that $ \procS \cap \procT = \emptyset $ and thus $ \procS \uplus \procT = \procS \cup \procT $.
We say that a condition $ \pred : \left( \procS \uplus \procT \right) \to \bool $ is \emph{preserved} by an encoding if for all source terms $ S $ that satisfy $ \pred $, the condition $ \pred $ also holds for $ \Encoding{S} $. A condition is \emph{reflected} by an encoding if whenever $ \Encoding{S} $ satisfies it, then so does $ S $.
Finally an encoding \emph{respects} a condition if it both preserves and reflects it.


\section{Encodability Criteria}
\label{sec:encodabilityCriteria}

Encodings are used to compare process calculi and to reason about their expressive power. Encodability criteria are conditions that limit the existence of encodings. Their main purpose is to rule out trivial or meaningless encodings, but they can also be used to limit attention to encodings that are of special interest in a particular domain or for a particular purpose.
These quality criteria are the main tool in \emph{separation results}, saying that one calculus is not expressible in another one; here one has to show that no encoding meeting these criteria exists. To obtain stronger separation results, care has to be taken in selecting quality criteria that are not too restrictive.  For \emph{encodability results}, saying that one calculus is expressible in another one, all one needs is an encoding, together with criteria testifying for the
quality of the encoding. Here it is important that the criteria are not too weak.

In the literature various different criteria and different variants of the same criteria are employed to achieve separation and encodability results \cite{vG94a,nestmann00,nestmannPierce00,palamidessi03,PalamidessiSVV06, nestmann06,vigliottiPhillipsPalamidessi07,CCAV08,parrow08,BusiGZ09,gorla10,petersNestmann12,vG12}.
Some criteria, like full abstraction or operational correspondence, are used frequently.
Other criteria are used to enforce a property of encodings that might only be necessary within a certain domain.
For instance, the homomorphic translation of the parallel operator---in general a rather strict criterion---was used in \cite{palamidessi03} to show the absence of an encoding from the synchronous into the asynchronous \piCal, because this requirement forbids for the introduction of global coordinators. Thus this criterion is useful when reasoning about the concurrent behaviour of processes, although it is in general too strict to reason about their interleaving behaviour. Unfortunately, it is not always obvious or clear whether the criteria used to obtain a result in a particular setting do indeed fit to this setting. Indeed, as discussed in \cite{petersNestmann12}, the homomorphic translation of the parallel operator forbids more than global coordinators, \ie is too strict even in a concurrent setting (compare to Section~\ref{sec:domainSpecificCriteria}).

In the following we shortly revise some of the most commonly used encodability criteria. In \cite{nestmann96} also a class of rather quantitative criteria for the effectiveness or efficiency of an encoding is discussed. The efficiency of an encoding can be measured for example by a criterion to count the number of messages or steps of an emulation \cite{buckleySilberschatz83, arunkummarHennessy92, knabe93}. However, here we are more focused on semantic and structural criteria.

\subsection{Direct Comparison via a Relation}

The least debatable criterion is the direct comparison of the source and the target language by a behavioural equivalence (or preorder). This criterion requires that $ \forall S \in \procS \logdot \Encoding{S} \mathcal{R} S $ holds, for some behavioural equivalence (or preorder) $ \mathcal{R} \; \subseteq \left( \procS \uplus \procT \right) \times \left( \procS \uplus \procT \right) $ that is either defined in exactly the same way on the source and the target language (as \eg in \cite{parrow08}) or explicitly distinguishes between source and target terms (as \eg in \cite{petersGlabbeek15}). Intuitively, it is required that the encoding respects the semantics of the source modulo the chosen relation. Obviously, the quality of the encoding directly depends on the choice of $ \mathcal{R} $. A stricter such relation leads to stricter requirements on the encoding function. It is also very clear in this case under which circumstances two different results can be compared. If both results are proven with respect to the same relation then the results can be compared directly. If one of the considered relations is strictly weaker then the results can be compared with respect to the weaker relation. Else, if the relations are incomparable, also the results are incomparable. Hence, a language $ \lang_1 $ can be considered as strictly weaker than the language $ \lang_2 $ with respect to $ \mathcal{R} $, if there is an encoding from $ \lang_1 $ into $ \lang_2 $ that satisfies the above requirement but there is no such encoding from $ \lang_2 $ into $ \lang_1 $.

Of course, there may be still some debate on how to choose the relation $ \mathcal{R} $. For instance neither the identity nor $ \mathcal{R} \; = \left( \procS \uplus \procT \right) \times \left( \procS \uplus \procT \right) $ are intuitively meaningful choices. Moreover, as shown in \cite{vanglabbeek01, vanglabbeek93} there are usually very many potential candidates. Variants of bisimulation are usually a rather strong candidate suitable for encodability results, whereas weaker candidates such as trace equivalence are better suited for separation results.
For a congruence it is sometimes necessary to restrict the set of considered contexts to contexts that respect the protocol behind the encoding (see \eg \cite{victorParrow96, glabbeek18}).
However, since the choice of the equivalence directly monitors the requirements on the encoding function, this problem is not that serious. There are, however, two serious drawbacks of this criterion. The first is the requirement that $ \mathcal{R} $ is either defined in exactly the same way on the source and the target language or explicitly distinguishes between source and target terms.
Usually behavioural relations are designed for one specific language, so it is not clear how to define a suitable relation if the two languages have no standard behavioural equivalence (or preorder) in common. A standard candidate for such an equivalence is weak reduction bisimulation, that is defined similarly on all calculi. Unfortunately, the variant of weak bisimulation given in Definition~\ref{def:bisimulation} is trivial.
We need to add a condition that compares the observables of bisimilar states in order to obtain a meaningful relation. But, since the source and target language might specify different standard observables, this equivalence (if it can be constructed at all) will usually be very complex and unreadable.
Note that \cite{gorla10} solves this problem by using a special observable called success instead of the standard observables of the languages, whereas \cite{fu16, glabbeek18} use relations to compare observables of different languages.
As a consequence, if $ \mathcal{R} $ is not a standard equivalence, the direct comparison of source and target terms modulo $ \mathcal{R} $ reveals less intuition on the encoding function. The second problem is that a complex $ \mathcal{R} $ leads to a hard proof of this criterion. If $ \mathcal{R} $ is not a standard equivalence, most of the standard techniques that would ease such a proof may not be applicable.

\subsection{Full Abstraction}
\label{sec:fullAbstraction}

Whenever source and target can not be compared directly with respect to a standard equivalence, full abstraction might be a way to nonetheless use standard equivalences.
\emph{Full abstraction}---denoted as \emph{observational correspondence} in \cite{fuLu10}---is probably the most common quality criterion for language comparison. It is used for instance in \cite{riecke91, mitchell93, sangiorgi94, yoshida96, nestmannPierce00, baldamusParrowVictor05, perez09}, just to name some. Full abstraction as proof method for language comparison was adapted from the use of full abstraction to show correspondence between a denotational semantics of a program and its operational semantics. An encoding $ \encoding $ is fully abstract if
\begin{align*}
	\forall S_1, S_2 \in \procS \logdot S_1 \RelS S_2 \quad \text{ iff } \quad \Encoding{S_1} \RelT \Encoding{S_2}
\end{align*}
for two behavioural equivalences $ \RelS \; \subseteq \procS \times \procS $ and $ \RelT \; \subseteq \procT \times \procT $, \ie full abstraction requires that equivalent source terms have to be mapped into equivalent target terms and vice versa. Note that the direction from the left to the right is often called soundness condition and the only if part completeness condition of full abstraction. The soundness condition is usually the most demanding part. Note that some well-known and widely accepted encodings, as \eg \cite{boudol92, hondaTokoro91, milner92, milner93}, do not satisfy this property with respect to a reasonable combination of standard equivalences. The main advantage of full abstraction is its wide applicability also with respect to (more or less) standard equivalences. It does \eg not require that source and target share any notion of observable, which is a premise for the use of most of the standard equivalences in the criterion above. However, again there may be a very large number of equivalences on the source as well as equivalences on the target and the strictness of the property expressed by full abstraction strongly relies on the combination of the chosen equivalences. To reduce the strong dependence of full abstraction results on the chosen equivalences, full abstraction is often combined with operational correspondence. In \cite{fuLu10} it is even stated that full abstraction is not of much use without operational correspondence. The two papers \cite{gorla2014abstraction, parrow2014abstraction} come to a similar conclusion after discussing potential pitfalls and misunderstandings \wrt full abstraction. These papers also hint to earlier such discussions. The possibility to chose a combination of standard equivalence that turn full abstraction trivial is a major drawback of this criterion. Another major drawback is that, because of the various possibilities to choose these two equivalences, it is often not possible to compare different full abstraction results.

\subsection{Operational Correspondence}

Intuitively, \emph{operational correspondence} requires executions to be respected. Again, it consists of a completeness and a soundness part. The completeness condition, also called adequacy, requires that for all source term steps $ S \stepS S' $ or source term executions $ S \stepsS S' $ there is one emulation in the target language such that $ \Encoding{S} \stepsT \relT \Encoding{S'} $, where $ \relT \; \subseteq \procT \times \procT $ is some equivalence on the target language. Note that there is no difference in the consideration of single source term steps or source term executions. Intuitively, the completeness condition requires that any source term execution is emulated by the target term modulo some equivalence $ \relT $. Again, completeness is usually the easier part.

For the soundness condition we basically find two formulations. The stricter formulation requires that for all executions of the target $ \Encoding{S} \stepsT T $ there exists some execution of the source $ S \stepsS S' $ such that $ \Encoding{S'} \relT T $.
Intuitively, soundness requires that whatever $ \Encoding{S} $ can do is a translation of some behaviour of $ S $ modulo $ \relT $. The weaker formulation requires that for all executions of the target $ \Encoding{S} \stepsT T $ there exists some execution of the source $ S \stepsS S' $ and some execution of the target $ T \stepsT T' $ such that $ \Encoding{S'} \relT T' $. Intuitively, it states that any execution of the target is some part of the emulation of an execution in the source modulo $ \relT $ \cite{parrowSjoedin92, gorla10}.
The main difference is that the latter formulation allows for intermediate or partial commitment states \cite{parrowSjoedin92, peters12, cspToCcs15}, \ie for states that do not need to be related directly to the states of the respective source term but that have to belong to some emulation of a source term step. In this sense, an intermediate state results from the partial emulation of a source term step.
In particular the last variant, proposed in \cite{gorla10}, was used for numerous encodability and separation results.

\begin{definition}[Operational Correspondence]
	\label{def:operationalCorrespondence}
	An encoding $ \encoding: \procS \to \procT $ is \emph{strongly operationally corresponding} \wrt $ \relT \subseteq \procT^2 $ if it is:
	\begin{compactitem}
		\item[\quad Strongly Complete:] $ \forall S, S' \logdot S \stepS S' \text{ implies } \left( \exists T \logdot \Encoding{S} \stepT T \wedge \left( \Encoding{S'}, T \right) \in \relT \right) $
		\item[\quad Strongly Sound:] $ \forall S, T \logdot \Encoding{S} \stepT T \text{ implies } \left( \exists S' \logdot S \stepS S' \wedge \left( \Encoding{S'}, T \right) \in \relT \right) $
	\end{compactitem}
	$ \encoding: \procS \to \procT $ is \emph{operationally corresponding} \wrt $ \relT \subseteq \procT^2 $ if it is:
	\begin{compactitem}
		\item[\quad Complete:] $ \forall S, S' \logdot S \stepsS S' \text{ implies } \left( \exists T \logdot \Encoding{S} \stepsT T \wedge \left( \Encoding{S'}, T \right) \in \relT \right) $
		\item[\quad Sound:] $ \forall S, T \logdot \Encoding{S} \stepsT T \text{ implies } \left( \exists S' \logdot S \stepsS S' \wedge \left( \Encoding{S'}, T \right) \in \relT \right) $
	\end{compactitem}
        $ \encoding: \procS \to \procT $ is \emph{weakly operationally corresponding} \wrt $ \relT \subseteq \procT^2 $ if it is:
	\begin{compactitem}
		\item[\quad Complete:] $ \forall S, S' \logdot S \stepsS S' \text{ implies } \left( \exists T \logdot \Encoding{S} \stepsT T \wedge \left( \Encoding{S'}, T \right) \in \relT \right) $
		\item[\quad Weakly Sound:] $ \forall S, T \logdot \Encoding{S} \stepsT T \text{ implies } \left( \exists S', T' \logdot S \stepsS S' \wedge T \stepsT T' \wedge \left( \Encoding{S'}, T' \right) \in \relT \right) $
	\end{compactitem}
\end{definition}

\noindent
Again different variants of operational correspondence may arise from different requirements on the assumed equivalence $ \relT $ on the target language and a wrong choice might turn operational correspondence trivial. In particular, each encoding is operational corresponding \wrt the universal relation on target terms.

In \cite{fuLu10} a version of operational correspondence is presented, that considers labelled steps instead of a reduction semantics under the assumption that there exists a mapping $ \widehat{\cdot} $ from the labels of the source term into the labels of the target term. Hence, the resulting requirement---$ \Encoding{S} \LabelledSteps{\widehat{\lambda}}_\indexTarget \relT \Encoding{S'} $ whenever $ S \LabelledSteps{\lambda}_\indexSource S' $ and $ \Encoding{S} \LabelledSteps{\lambda}_\indexTarget T $ implies $ S \LabelledSteps{\lambda'}_\indexSource S' $ for some $ \lambda', S' $ such that $ \Encoding{S'} \relT T $ and $ \widehat{\lambda'} = \lambda $---can be considered as stricter than the above variant of operational correspondence, because also observables have to be respected modulo $ \widehat{\cdot} $. In fact, without this strengthening to labelled semantics, operational correspondence alone can hardly be considered as suitable criterion. Hence, the above version based on reduction semantics is usually combined with other criteria as full abstraction or some requirements on the preservation or reflection of some kind of observable.

\subsection{Observables, Testing, and Termination}

If source and target terms can not be compared directly by a standard equivalence, \eg because not all standard observables of the source are standard observables of the target, a natural weaker requirement is to consider preservation or reflection of the remaining observables that are shared by source and target. Typical observables are links used for communication, barbs (communication capabilities), or traces \cite{vigliottiPhillipsPalamidessi07, parrow08}. Moreover, the use of \emph{termination properties} as the possibility of deadlock, livelock, or divergence are popular \cite{nestmann00, parrow08, gorla10, fuLu10}. Also all kind of \emph{tests} that the process may (or must) pass, for some formal notion of test can be used to compare source and target behaviour \cite{parrow08, gorla10}.

Another kind of termination property is the promptness condition. Intuitively, promptness ensures that an encoding does not introduce preprocessing steps.
An encoding $ \encoding $ is \emph{prompt}, if $ S \noStep_\indexSource $ implies $ \Encoding{S} \noStep_\indexTarget $ for all source terms $ S \in \procS $.

In \cite{gorla10} the criterion \emph{success sensitiveness} was proposed. An encoding is success sensitive if it respects reachability of a particular process $ \success $ that represents successful termination, or some other form of success, and is added to the syntax of the source as well as the target language. Since $ \success $ cannot be further reduced and $ \Names{\success} = \FreeNames{\success} = \BoundNames{\success} = \emptyset $, the semantics and structural congruence of a process calculus are not affected by this additional constant operator. We write $ P\hasSuccess $ to denote the fact that $ P $ is successful---however this predicate might be defined in the particular source or target language. Reachability of success is then defined as $ P\reachSuccess \deff \exists P' \logdot P \steps P' \wedge P'\hasSuccess $. We use may-testing here, but alternatively \eg must-testing or fair-testing can also be implemented. An encoding is success sensitive if each source term and its translation answer the test for reachability of success in the same way.

\begin{definition}[Success Sensitiveness]
	\label{def:successSensitiveness}
	Let $ \langS $ and $ \langT $ each define a predicate $ \cdot\hasSuccess : \proc \to \bool $.
	An encoding $ \encoding : \procS \to \procT $ is \emph{success sensitive} if, for all $ S \in \procS $, $ S\reachSuccess $ iff $ \Encoding{S}\reachSuccess $.
\end{definition}

Note that $ \success $ can be considered as some kind of termination property---describing successful termination in contrast to not successful termination by a term that cannot reduce but is not $ \success $---or as the successful pass of some kind of test.

If the source and target language share standard observables, we can easily extend success sensitiveness to barb sensitiveness. Assume that, instead of $ \cdot\hasSuccess : \proc \to \bool $, the languages $ \langS $ and $ \langT $ each define a predicate $ \HasBarb{\cdot}{\cdot} : \proc \times \mathcal{B} \to \bool $. An encoding $ \encoding : \procS \to \procT $ \emph{strongly respects barbs} if, for all $ S \in \procS $ and all $ \alpha \in \mathcal{B} $, $ \HasBarb{S}{\alpha} $ iff $ \HasBarb{\Encoding{S}}{\alpha} $. A weak variant is obtained by replacing the predicate $ \HasBarb{\cdot}{\cdot} $ for the existence of a barb by a predicate $ \ReachBarb{\cdot}{\cdot} $ that is true if a barb is reachable.

\subsection{Structural Requirements}
\label{sec:structuralRequirements}

The above discussed criteria describe semantic requirements, \ie requirements on the behaviour of target terms. To prove the quality of an encoding semantic criteria are often combined with structural criteria. Intuitively, the semantic criteria describe how the encoded terms should behave with respect to the behaviour of the corresponding source term, whereas structural criteria rather describe how the encoded terms have to look like. Moreover, as stated in \cite{parrow08}, structural criteria are needed in order to measure expressiveness of operators in contrast to expressiveness of terms.

The most common structural criterion is compositionality with homomorphy as a special case. Intuitively, \emph{compositionality} states that the translation of a compound term must be defined in terms of the translation of the subterms. To mediate between translations of subterms, a context is introduced. Different manifestations result from different requirements on allowed contexts. In the strictest form, often denoted as \emph{homomorphy}, the context has to be the original source term operator again, \ie an encoding translates the source term operator $ \opn\left( x_1, \ldots, x_n, S_1, \ldots, S_m \right) $ homomorphically if it ensures that $ \Encoding{\opn\left( x_1, \ldots, x_n, S_1, \ldots, S_m \right)} = \opn\left( x_1, \ldots, x_n, \Encoding{S_1}, \ldots, \Encoding{S_m} \right) $ holds for all $ x_1, \ldots, x_n \in \names $ and all $ S_1, \ldots, S_m \in \procS $.
Of course, homomorphy requires that the respective operator is part of the source and the target language. Because of this, homomorphy is often required only for specific common operators such as the parallel operator, because it occurs more or less with the same meaning and comparable syntax in most of the process calculi.
If we assume that the parallel operator is a binary operator in the source and the target language, the homomorphic translation of the parallel operator means $ \Encoding{S_1 \mid S_2} = \Encoding{S_1} \mid \Encoding{S_2} $ for all $ S_1, S_2 \in \procS $, where $ \mid $ is the syntactical representation of the parallel operator.
If single names are translated into single names, it makes sense to extend the notion of homomorphic translation to the use of a renaming policy. Thus, if the encoding translates an operator $ \opn\left( x_1, \ldots, x_n, S_1, \ldots, S_m \right) $ always into $ \opn\left( \phi\!\left(x_1\right), \ldots, \phi\!\left(x_n\right), \Encoding{S_1}, \ldots, \Encoding{S_m} \right) $, we call the translation of this operator again homomorphic.

Homomorphic translations of operators are \eg used to analyse the expressive power of a single operator. To show for instance that a set of operators is not minimal the existence of an encoding is analysed that translates all operators homomorphically except for the operator that should be removed. Moreover, homomorphy is a very nice property, because it significantly eases the proof of the correctness of the encoding function. Basically, the homomorphic translation of an operator ensures that for this operator nothing is to show, because in this point the encoding obviously preserves and reflects all properties of that operator. However, even in case the respective operator is part of the source as well as the target language, homomorphy is a very strict requirement. Intuitively, it states that the encoding function is not allowed to touch the respective operator and, hence, is not allowed to simulate its behaviour by some protocol. Such translations are possible only if the compared languages are very close (at least with respect to this operator).

\cite{palamidessi03, gorla10, peters12} show that not even between calculi that are so close as the full \piCal (with mixed choice) and its subcalculus with only separate choice an encoding that translates the parallel operator homomorphically exists. Instead, often compositionality is required.
Intuitively, an encoding is compositional if the translation of an operator is the same for all occurrences of that operator in a term. Hence, the translation of that operator can be captured by a context that is allowed in \cite{gorla10} to be parametrised on the free names of the respective source term.
\begin{definition}[Compositionality]
	\label{def:compositionality}
	The encoding $ \encoding $ is \emph{compositional} if, for every operator $ \opn : \names^n \times \procS^m \to \procS $ and for every subset of names $ N $, there exists a context $ \Context{N}{\opn}{\hole_1, \ldots , \hole_{n + m}} : \names^n \times \procS^m \to \procT $ such that, for all $ x_1, \ldots, x_n \in \names $ and all $ S_1, \ldots, S_m \in \procS $ with $ \FreeNames{S_1} \cup \ldots \cup \FreeNames{S_m} = N $, it holds that $ \Encoding{\opn\left( x_1, \ldots, x_n, S_1, \ldots, S_m \right)} = \Context{N}{\opn}{x_1, \ldots, x_n, \Encoding{S_1}, \ldots, \Encoding{S_m}} $.
\end{definition}
It does not impose additional restrictions on the introduced context. For many encodings the parameter $ N $ is not relevant or can be omitted by using instead a renaming policy on the source term names in the generated context. In contrast to homomorphy, compositionality is a very natural requirement. Intuitively, it states that every occurrence of an operator in the source term is treated by the encoding function in exactly the same way, \ie is translated into the same term modulo the translation of the respective subterms.
Also note that a compositional encoding, \ie an encoding that translates all source term operators compositionally, implies that also any source term context can be represented as a context in the target language \cite{parrow08}. Moreover, compositionality guides the design of encodings, because it describes how an encoding has to look like.

Another structural criterion is the \emph{preservation of substitutions}, denoted as \emph{name invariance} in \cite{gorla10} and as \emph{stability} in \cite{fuLu10}. It usually requires that, for all source terms $ S \in \procS $ and all substitutions $ \sigma $ on source terms, there exists some substitution $ \sigma' $ on target terms such that $ \Encoding{\sigma\!\left( S \right)} \asymp_T \sigma'\!\left( \Encoding{S} \right) $ for some equivalence $ \asymp_T \; \subseteq \procT \times \procT $ on target terms. Often additional requirements on the relationship between $ \sigma $ and $ \sigma' $ or on the equivalence $ \asymp_T $ are stated. The strictest case is of course that $ \sigma = \sigma' $ and $ \asymp_T \; = \; \equiv_\alpha $. This criterion is based on the idea that names are property-less \cite{fuLu10}. Hence, the preservation of substitutions should ensure that encodings of source terms that differ only in their free names can also only differ in free names (modulo the provided equivalence).

In \cite{gorla10} name invariance is defined modulo the introduced renaming policy. Accordingly, an encoding is considered independent of specific names if it preserves all substitutions $ \sigma $ on source terms by a substitution $ \sigma' $ on target terms such that $ \sigma' $ respects the changes made by the renaming policy.

\begin{definition}[Name Invariance]
	\label{def:nameInvariance}
	The encoding $ \encoding $ is \emph{name invariant} if, for every $ S \in \procS $ and $ \sigma $, it holds that 
	\begin{align*}
		\Encoding{\sigma\left( S \right)} \begin{cases} \equiv_\alpha \sigma'\left( \Encoding{S} \right) & \quad \text{if } \sigma \text{ is injective}\\ \relT \sigma'\left( \Encoding{S} \right) & \quad \text{otherwise} \end{cases}
	\end{align*}
	where $ \sigma' $ is such that $ \phi\!\left(\sigma\left( n \right)\right) = \sigma'\left( \phi(n) \right) $ for every $ n \in \names $.
\end{definition}

Moreover, \cite{vigliottiPhillipsPalamidessi07} present \emph{link independence}, a condition that prevents encodings from introducing free names. More precisely, link independence means that, for all source terms $ S_1, S_2 \in \procS $, $ \FreeNames{S_1} \cap \FreeNames{S_2} = \emptyset $ implies $ \FreeNames{\Encoding{S_1}} \cap \FreeNames{\Encoding{S_2}} = \emptyset $.

We denote the criteria presented in this subsection as \emph{structural criteria} because they focus mainly on structural properties of the encoding function. We should, however, be aware that every criterion limits the existence of encoding functions and that such limitations usually also have a semantic effect. The main purpose of compositionality is to ensure that the encoding function is compositional. But compositionality also forbids for global coordination (see Section~\ref{sec:domainSpecificCriteria}). Because of that, there is a number of well-accepted encodings---as \eg the encoding of the \piCal into the \joinCal in \cite{fournetGonthier96}---that are not compositional. Instead the encoding in \cite{fournetGonthier96} consists of two levels: an outer level that is parametrised on the free names of the source term and an inner compositional encoding.

\subsection{Domain-Specific Criteria}
\label{sec:domainSpecificCriteria}

Above we presented different kinds of criteria that were used to specify the notion of a ``good'' encoding in the literature. Thereby the purpose of all criteria introduced so far, is to define a general notion of correctness for encodings.
We know that there is no agreement about the choice, combination, or concrete variant of the above criteria for a general notion of correctness or quality of an encoding. But even if we would agree on such a general notion of quality, there would still be the need for domain-specific criteria.
A general notion is in particular necessary, if we want to compare different results or build a hierarchy. Accordingly, such a general notion of quality is a good starting point for language comparison. Nonetheless, we will sometimes need to extend it by domain-specific criteria. Domain-specific criteria, as the name suggests, are used to analyse properties of a specific domain that may in general not be interesting. Hence, it is not a good idea to overload a general framework by permanently adding domain-specific criteria. Instead, we add a domain-specific criterion only if it is necessary to answer a particular kind of question. Possible domains in the context of process calculi are \eg causality, the branching time behaviour of processes, or considerations related to some special features as failures or time constrains.

Note that an additional criterion may strengthen an encodability result, but it weakens separation results. Moreover, already a single additional criterion significantly complicates the comparison of a result with other already established results that do not rely on this additional criterion. Quite often, they even lead to incomparable results; in particular if two results use different additional criteria. Hence, domain-specific criteria should only be used if they are unavoidable to answer a specific kind of question.

As an example for a domain-specific criterion, we discuss how to obtain a criterion that ensures that an encoding preserves the degree of distribution. Given an extension of a process calculus with an explicit notion of distribution or location we can define the degree of distribution of a system as the number of its locations. If locations are not defined explicitly, a process $ P $ is distributable into $ P_1, \ldots, P_n $, if we find some distribution that extracts $ P_1, \ldots, P_n $ from within $ P $ onto different locations. Preservation of distribution then means that the target term is as least as distributable as the source term. Note that the operator of process calculi that is usually associated with distribution is the parallel operator. Accordingly, we consider components of a term that are composed in parallel as distributable. More precisely, we understand distribution as the separation of a process into its (sequential) components.

Hence, by studying distribution preserving encodings, we analyse the possibilities to implement the operators of a calculus or especially its parallel operator. If it is always possible to preserve the degree of distribution in an encoding of a source language into a target language which is close to an implementation \eg in a real world scenario, then the corresponding parallel operator can be implemented in this scenario simply as the operator of distribution, \ie parallel source terms can be implemented in distributed real world processes. If it is not possible to obtain a distribution preserving encoding, then the source language implicitly defines side conditions on the use of the parallel operator usually induced by the defined synchronisation mechanism that forbids for such simple implementations. Thus, the implementation of parallel source terms as distributed processes may be possible only under some side conditions, which are hopefully already paraphrased by the respective separation result. In particular, an encoding that preserves the degree of distribution should not introduce a coordinator for concurrent actions, because if concurrent actions are coordinated by the same instance they are sequentialised, \ie the implementation is less efficient.

Note that compositionality in Definition~\ref{def:compositionality} already prevents from the use of global coordinators. Compositionality requires that all occurrences of a parallel operator have to be translated basically in the same way. Hence, if such an encoding introduces a coordinator then for every parallel operator a coordinator is introduced and there is no possibility to examine which of them is the outermost or to order them, \ie it is not possible to coordinate the coordinators such that they proceed as a centralised entity. In that view, compositionality can be seen as a minimal criterion to ensure the preservation of distribution. However, compositionality alone is too weak, because it still allows for \emph{local coordinators}, \ie a compositional encoding may still sequentialise some parts of a source term (see \eg the encodings in \cite{petersNestmann12, cspToCcs15}).

Instead the homomorphic translation of the parallel operator, \ie $ \Encoding{P \mid Q} = \Encoding{P} \mid \Encoding{Q} $, was often used as a criterion to measure whether an encoding respects the degree of distribution (see \eg \cite{palamidessi03, carboneMaffeis03, laneveVitale10}). The homomorphic translation of the parallel operator forbids the introduction of (global and local) coordinators for the translation of the parallel operator. As discussed in \cite{peters12, petersNestmann12, petersNestmannGoltz13} the homomorphic translation of the parallel operator usually implies that the respective encoding indeed preserves the degree of distribution but that the converse is not true, \ie there are encodings that do not translate the parallel operator homomorphically but preserve nonetheless the degree of distribution of all source terms. In this sense, the homomorphic translation of the parallel operator is too strict---at least for separation results. It rightly forbids the introduction of coordinators that reduce the degree of distribution. But it also forbids protocols that handle communications of parallel components without sequentialising them or reducing the degree of distribution in another sense. Moreover, the homomorphic translation of the parallel operator is not always suited to reason about distribution in process calculi. For example in the \joinCal it is not always possible to separate distributed components by means of a parallel operator.

\cite{peters12, petersNestmann12, petersNestmannGoltz13} introduce an alternative criterion for the preservation of the degree of distribution. Compositionality and the homomorphic translation of the parallel operator are both structural criteria. Hence, one may assume that also the preservation of distribution is a structural criterion, but in fact this is not true. A natural first condition is to require that encoded source terms are at least as distributable as the source term itself, \ie that the degree of distribution has to be preserved by the encoding. However, it does not suffice to reason about the degree of distribution, \ie about the number of distributable components, without additional requirements on the components. An encoding can always trivially ensure that the encoding has at least as much distributable components by introducing new components without any behaviour. Thus, we require that the encodings of distributable source term parts and their corresponding parts in the encoding are related by $ \relT $. By doing so we relate the definition of the preservation of distributability to operational completeness, \ie a semantic criterion that ensures the preservation of the behaviour of the source term (part). Hence, we require that each target term part can emulate at least all behaviour of the respective source part.
As a side effect, we require, that whenever a part of a source term can solve a task independently of the other parts, \ie it can reduce on its own, then the respective part of its encoding must also be able to emulate this reduction independently of the rest of the encoded term. This reflects our intuition that distribution adds some additional requirements on the independence of parallel terms. Accordingly, we require that not only the source term and its encoding are distributable to the same degree, but also their derivatives, \ie we do not only consider the \emph{initial} degree of distribution. Because of that, the criterion that is presented in \cite{petersNestmannGoltz13} has both a structural as well as a semantic component.
Remember that a term $ P $ is distributable into $ P_1, \ldots, P_n $, if we find some distribution that extracts $ P_1, \ldots, P_n $ from within $ P $ onto different locations.

\begin{definition}[Preservation of Distribution]
	\label{def:distributabilityPreservation}
	An encoding $ \encoding : \procS \to \procT $ \emph{preserves distribution} if for every $ S \in \procS $ and for all terms $ S_1, \ldots, S_n \in \procS $ that are distributable within $ S $ there are some $ T_1, \ldots, T_n \in \procT $ that are distributable within $ \Encoding{S} $ such that $ T_i \relT \Encoding{S_i} $ for all $ 1 \leq i \leq n $.
\end{definition}

In essence, this requirement is a distributability-enhanced adaptation of operational completeness. Whenever a source term is distributable into $ n $ terms then its encoding must again be distributable into $ n $ terms, \ie the encoded source term is at least as distributable as the source term itself. Moreover, if some of these $ n $ terms, say $ S_i $, can perform some execution independently of the rest then, by operational completeness, this execution has to be emulated by its translation modulo $ \relT $, \ie $ S_i \stepsS S_i' $ implies $ T_i \stepsT \relT \Encoding{S_i'} $. This formalisation of the preservation of distributability respects both the intuition on distribution as separation on different locations---captured by the structural requirement that the encoded source term is at least as distributable as the source term itself---as well as the intuition on distribution as independence of processes and their executions---a semantic requirement implemented by the condition $ T_i \relT \Encoding{S_i} $.

The main disadvantage of this criterion is its complexity. It is more model-independent, since it does not rely on the notion of the parallel operator, \ie is applicable also for calculi without a parallel operator or completely different kinds of parallel operators such as in \csp and it can be applied to calculi like the \joinCal, where distributed components are not always separated by a parallel operator. But, in order to apply this criterion, we have to specify in the source and the target language what it means for a process to be distributable into a set of components. In calculi with explicit locations this is usually easy. But it is difficult to find a general, \ie model-independent, formalisation of distributability (see \cite{peters12, petersNestmann12, petersNestmannGoltz13}). Moreover, this criterion was designed in order to be combined with operational correspondence and is strongly connected to this criterion.
The advantage of the homomorphic translation of the parallel operator is that it is simple, easy to understand, and described independently of other criteria. But it is too strong for separation results, whereas compositionality is too weak for encodability results.

\subsection{Summary}

We introduced a number of encodability criteria. We observe that these criteria appear in different variants and that there is no agreement on a set of criteria that should be used; not even if we ignore domain-specific criteria. Moreover, we observe that some criteria are often used but as the discussion on full abstraction (see Section~\ref{sec:fullAbstraction} and \cite{fuLu10, gorla2014abstraction, parrow2014abstraction}) shows not always well-understood. For domain-specific criteria the situation is usually worse, \ie it is even more difficult to analyse whether they are suitable. A mechanism to reason about the quality of encodability criteria is discussed in Section~\ref{sec:evaluation}.

Encodability criteria define properties of the encoding function and should ensure that an encoding can be considered as meaningful. Unfortunately, encodability and separation results that are derived \wrt different sets of encodability criteria or different variants of these criteria are hard to compare. Section~\ref{sec:frameworks} analyses general frameworks of encodability criteria. Note that the formalisation of a criterion should be as general as possible, because a formalisation that is to close to a specific process calculus may hinder the derivation of similar results for other calculi and, thus, the comparison with other results. Furthermore, it is sometimes easier to define a new criterion with respect to an existing one, but again this may shrink the possibilities to compare to other results that do not satisfy the old criterion.

Finally, note that the criteria pose different kinds of proof obligations on the correctness proofs of encodings. Structural criteria (and also the criteria for the efficiency of an encoding function) are usually easier to verify than the remaining semantic criteria. But the semantic criteria are often considered as more essential.
Unfortunately, we are not only lacking mechanisms to analyse the quality of encodability criteria and an agreement of a general notion of a ``good'' encoding but also general proof techniques for encodability criteria. From the criteria introduced above full abstraction and operational correspondence are usually the most elaborate criteria to prove. In the case of full abstraction the proofs heavily depend on the chosen pair of equivalences.
Operational correspondence is usually shown by induction on the nature of source or target term steps. But some more sophisticated study of proof techniques might provide \eg some hints on how to deal with the intermediate states that weakly operationally corresponding encodings might introduce (compare \eg to the discussion of pre- and post-processing steps in \cite{peters12}).


\section{A General Notion of Quality}
\label{sec:frameworks}

A general notion of quality is important to reason about the general expressive power of languages and to compare different results. In particular, they are essential to build a hierarchy that can only be based on encodability and separation results \wrt the same set of criteria. Moreover, if we want to relate the expressive power of two given languages there is no guidance on how to start or whether an obtained result is sufficiently substantiated by the chosen criteria to call it reasonable. There are basically two problems that we have to face in providing such a general notion:
\begin{inparaenum}[(1)]
	\item We have to choose the nature of criteria. For a general setting the criteria should be model-independent. This also includes to some extent the relations that are used \eg in full abstraction and operational correspondence, \ie a general setting has to specify how to choose them. Moreover, the criteria should be designed to capture properties of the encoding that are generally agreed as being useful.
	\item We have to decide on the variants of the respective criteria. Weaker criteria allow for more encodability and less separation results, whereas more restrictive criteria allow for less encodability and more separation results. The difficulty is to identify the sweet spot between too restrictive and too weak criteria such that the variants are meaningful for separation as well as encodability results.
\end{inparaenum}
As discussed in Section~\ref{sec:domainSpecificCriteria}, there is a need for domain-specific criteria. Accordingly, it should be possible to extend a general framework by domain-specific criteria.

\subsection{Towards A Unified Approach to Encodability and Separation Results}

In order to provide a general framework, \cite{gorla08, gorla10} suggests five criteria well suited for language comparison, \ie for positive as well as negative translational results. As claimed in \cite{gorla10}, most of the encodings appearing in the literature satisfy this framework and several known separation results can also be derived within this framework but there are also encodings that do not satisfies this framework, \ie the framework is not trivial. The set of criteria is small and handy but at the same time guides the design of encoding functions and supports the proof of translational results by separating the requirements on different intuitive criteria. This framework specifies an encoding to be ``good'' if it satisfies the five presented criteria. In \cite{gorla08, gorla09, gorla10DistComp, gorla10} a number of encodability and separation results---including some new results---that are derived in this setting can be found.

The five conditions are divided into two structural and three semantic criteria. The structural criteria include
\begin{inparaenum}[(1)]
	\item \emph{compositionality} and
	\item \emph{name invariance}. The semantic criteria include
	\item \emph{operational correspondence},
	\item \emph{divergence reflection}, and
	\item \emph{success sensitiveness}.
\end{inparaenum}
Note that for the definition of name invariance and operational correspondence a behavioural equivalence $ \relT $ for the target language is assumed. Its purpose is to describe the abstract behaviour of a target process, where abstract refers to the behaviour of the source term.

The two structural criteria were introduced and discussed in Section~\ref{sec:structuralRequirements}. Compositionality is given in Definition~\ref{def:compositionality} and name invariance in Definition~\ref{def:nameInvariance}. They state that the encoding function should be compositional and independent of specific names.
To ensure that there are no conflicts between names that are introduced by the encoding function for technical reasons, \ie to implement some kind of protocol, and the source term names, the encoding is equipped with a renaming policy (Definition~\ref{def:renamingPolicy}).

The first semantic criterion and usually the most elaborate one to prove is weak operational correspondence as given in Definition~\ref{def:operationalCorrespondence}. The definition of operational correspondence relies on the equivalence $ \relT $ to get rid of junks possibly left over within computations of target terms.
To deal with infinite computations in encodings, the second semantic criterion requires that the encoding reflects divergence. It ensures that the encoding function cannot introduce new divergent behaviour, \ie all divergent target terms are due to the encoding of a divergent source term.
The last criterion links the behaviour of source terms to the behaviour of their encodings. \cite{gorla10} assumes a \emph{success} operator $ \success $ as part of the syntax of both the source and the target language. An encoding preserves the abstract behaviour of the source term if it and its encoding answer the tests for success in exactly the same way (Definition~\ref{def:successSensitiveness}). 
This criterion only links the behaviours of source terms and their literal translations, but not of their derivatives. To do so, \cite{gorla10} relates success sensitiveness and operational correspondence by requiring that the equivalence on the target language never relates two processes with different success behaviours.
\begin{definition}[Success Respecting]
	An equivalence $ \mathcal{R} $ is \emph{success respecting} if, for every $ P, Q \in \proc_C $ with $ P \reachSuccess $ and $ Q \not\reachSuccess $, it holds that $ \left( P, Q \right) \notin \mathcal{R} $.
	We require that $ \relT $ is a success respecting equivalence.
\end{definition}
The combination of success sensitiveness and operational correspondence allows to compare the behaviour of source and target terms even if they do not share common observables.
By \cite{gorla10} a ``good'' equivalence $ \relT $ is often defined in the form of a barbed equivalence (as described \eg in \cite{milnerSangiorgi92}) or can be derived directly from the reduction semantics (as described \eg in \cite{hondaYoshida95}) and is often a congruence, at least with respect to parallel composition.

\subsection{Theory of Interaction}

In contrast to \cite{gorla10}, the general framework in \cite{fuLu10} is based on labelled semantics. Intuitively, they combine full abstraction and operational correspondence into a bisimulation-like relation called \emph{subbisimulation}. Subbisimulation in \cite{fuLu10} connects labelled executions of the source and the target language. Moreover, preservation and reflection of divergence is required. The approach is then used to compare different variants of \ccs and different variants of the \piCal. For example subbisimilarity is applied to show the independence of the operators of the \piCal.

The paper \cite{fu16} extends this approach into a general theory of interaction. The prime motivation for the theory of interaction is to bridge the gap between the computation theory, \ie what kind of functions can be computed, and the interaction theory. Therefore four fundamental and model-independent principles are presented and from them a general theory of equality and expressiveness are derived. Again subbisimilarity but without labels is used as criterion for encodings.

\begin{definition}[Subbisimilarity]
	A relation $ \mathcal{R} $ is a \emph{subbisimilarity} if it is total, sound, equipollent, extensional, codivergent, and bisimilar.
\end{definition}

Intuitively,
\begin{inparaenum}[(1)]
	\item total means that for every source term there is a target term,
	\item soundness is similar to operational soundness,
	\item equipollence implies that related terms either both cannot reach any state with an observable or both can reach a state with (potentially different) observables,
	\item extensional means congruent \wrt the parallel operator and restriction,
	\item codivergent means that the relation is divergence respecting, and
	\item bisimilarity is defined similar to Definition~\ref{def:bisimulation}.
\end{inparaenum}

In comparison to the framework of Gorla the requirements induced by the formulation of subbisimilarity seem to be stricter, although a direct comparison is difficult, because of the different formulations. However, \cite{fuLu10} and \cite{fu16} make use of a stricter variant of operational correspondence that does not allow for intermediate or partially committed states. Moreover, \cite{fuLu10, fu16} fix a number of assumptions on process calculi, \eg that all process calculi contain at least the parallel operator and restriction.

\subsection{Musings on Encodings and Expressiveness}

Also \cite{vG12, glabbeek18} aims at providing a general notion of expressiveness for language comparison. Here an encoding should be \emph{valid} and \emph{correct}. In \cite{glabbeek18} these concepts are defined up to a semantic equivalence or preorder $ \sim $, that is not fixed by the framework.

Intuitively, an encoding is valid if it preserves the meaning of expressions, \ie such that the meaning of a
translated expression is semantically equivalent to the meaning of the original.
Therefore, \cite{glabbeek18} assumes a function for each language that associates the \emph{meaning} to processes by mapping them on values (for simplicity and in contrast to \cite{glabbeek18} I ignore in the following that processes may contain variables). Different languages may have different domains of values. The semantic relation $ \sim $ is used to mediate between these different domains, \ie it is assumed that it is a relation on values over a universe that contains the domains of the source as well as the target language. Moreover, \cite{glabbeek18} requires a \emph{semantic translation} $ \mathbf{R} $ that is a relation that relates each value of the source with its counterpart (or counterparts) in the target. Then, an encoding is \emph{correct} iff the meaning of the translation of an expression is a counterpart of the meaning of this expression.

\begin{definition}[Correctness]	
	An encoding $ \encoding $ is \emph{correct} \wrt a semantic translation $ \mathbf{R} $ if $ \mathbf{R} $ relates the meaning of $ S $ and $ \Encoding{S} $ for all $ S \in \procS $.

	An encoding $ \encoding $ is \emph{correct} up to $ \sim $ iff $ \sim $ is an
equivalence, the restriction $ \mathbf{R} $ of $ \sim $ to the cross product of the domain of the source and the target is a semantic translation, and $ \encoding $ is correct \wrt $ \mathbf{R} $.
\end{definition}

\begin{definition}[Validity]
	An encoding $ \encoding $ is \emph{valid} up to $ \sim $ iff it is correct \wrt some semantic translation $ \mathbf{R} \subseteq {\sim} $. Language $ \langT $ is at least as expressive as $ \langS $ up to $ \sim $ if an encoding valid up to $ \sim $ from $ \langS $ into $ \langT $ exists.
\end{definition}

As discussed in \cite{glabbeek18}, in comparison to \cite{gorla10} the above approach implies a slightly stricter variant of compositionality (without the parameter on a set of names $ N $) but no name invariance criterion. Moreover, the combination of three semantic criteria of \cite{gorla10} are close to an instantiation of the criteria in this approach with a particular preorder $ \sim $, namely a success respecting and divergence respecting variant of bisimulation (see Section~\ref{sec:evaluation}).


\section{Formalising and Analysing Encodability Criteria}
\label{sec:evaluation}

There exists a bunch of different criteria and different variants of criteria in order to reason in different settings. This leads to incomparable results. Moreover it is not always clear whether the criteria used to obtain a result in a particular setting do indeed fit to this setting.
A way to formally reason about and compare encodability criteria is presented in \cite{petersGlabbeek15}. The main idea of this approach is to map encodability criteria on requirements on a relation between source and target terms that is induced by the encoding function. This way the problem of analysing or comparing encodability criteria is reduced to the better understood problem of comparing relations on processes. An Isabelle/HOL formalisation can be found in \cite{archiveOfFormalProofs15}.

The different purposes of encodability criteria lead to very different kinds of conditions that are usually hard to analyse and compare directly. In fact even widely used criteria---as full abstraction---seem not to be fully understood by the community, as the need for articles as \cite{gorla2014abstraction, parrow2014abstraction} shows. In contrast to that, relations on processes---such as simulations and bisimulations---are a very well studied and understood topic (see for example \cite{vanglabbeek01}). Moreover, it is natural to describe the behaviour of terms, or compare them, modulo some equivalence relation. Also many encodability criteria, like operational correspondence, are obviously designed with a particular kind of relation between processes in mind. Therefore, in order to be able to formally reason about encodability criteria, to completely capture and describe their semantic effect, and to analyse side conditions of combinations of criteria, mapping them on conditions on relations between source and target terms seems to be natural.

According to \cite{petersGlabbeek15} every encoding $ \encoding : \procS \to \procT $ induces a relation $ \IRel \subseteq \left( \procS \uplus \procT \right)^2 $ on the disjoint union of its source and target terms by relating source terms and their literal translations, \ie $ \left( S, \Encoding{S} \right) \in \IRel $ for all $ S \in \procS $. Encodability criteria induce further properties on such relations. By analysing the different kinds of such properties the semantic effect of an encodability criterion can be studied and different criteria can be compared in a model-independent and formal way.
In order to completely capture the effect of a criterion, \cite{petersGlabbeek15} aims at iff-results of the form
\begin{quote}
	$ \encoding $ satisfies {\bf C} iff there exists a relation $ \IRel $ such that $ \forall S\logdot \left( S, \Encoding{S} \right) \in \IRel $ and $ \Pred{\IRel} $,
\end{quote}
where $ \pred $ is the condition that captures the effect of {\bf C}.
For example, an encoding reflects divergence iff there exists a relation $ \IRel $ such that $ \forall S\logdot \left( S, \Encoding{S} \right) \in \IRel $ and $ \IRel $ reflects divergence. Similarly, an encoding (weakly/strongly) respects barbs iff there exists a relation $ \IRel $ such that $ \forall S\logdot \left( S, \Encoding{S} \right) \in \IRel $ and $ \IRel $ (weakly/strongly) respects barbs.

Using this technique, \cite{petersGlabbeek15} shows that without further requirements on the source and target relations $ \RelS $ and $ \RelT $, that are used in the formulation of full abstraction, the semantic effect of full abstraction is very small. Full abstraction is mapped on a relation that relates at least each source term to its literal translation and includes the relations $ \RelS $ and $ \RelT $. Let us additionally add pairs of the form $ \left( \Encoding{S}, S \right) $ for all $ S \in \procS $.
Then, an encoding is fully abstract \wrt the preorders $ \RelS $ and $ \RelT $ iff there exists a transitive relation $ \IRel $ that relates at least each source term to its literal translation in both directions, such that the restriction of $ \IRel $ to source terms $ \Red{\IRel}{\procS} $ is $ \RelS $ and $ \Red{\IRel}{\procT} = \RelT $.

\begin{lemma}[Full Abstraction, \cite{petersGlabbeek15}]
	$ \encoding : \procS \to \procT $ is \emph{fully abstract} \wrt the preorders $ \RelS \subseteq \procS^2 $ and $ \RelT \subseteq \procT^2 $ iff $ \exists \IRel \logdot \left( \forall S \logdot \left( S, \Encoding{S} \right), \left( \Encoding{S}, S \right) \in \IRel \right) \wedge \RelS = \Red{\IRel}{\procS} \wedge \RelT = \Red{\IRel}{\procT} \wedge \IRel $ is transitive.
\end{lemma}

\noindent
Thus an encoding is fully abstract \wrt $ \RelS $ and $ \RelT $ if the encoding function combines the relations $ \RelS $ and $ \RelT $ in a transitive way. This underpins the discussions in \cite{fuLu10, gorla2014abstraction, parrow2014abstraction} showing that full abstraction without a clarification of the respective equivalences is not meaningful. It also shows the need for a formal analysis of encodability criteria.

The formulation of operational correspondence (in all its variants) strongly reminds us of simulation relations on processes, such as bisimilarity. Obviously this criterion is designed in order to establish a simulation-like relation between source and target terms. By mapping this criterion on properties of a relation, we can determine the exact nature of this relation.
The first two variants of Definition~\ref{def:operationalCorrespondence} exactly describe strong and weak bisimilarity up to $ \relT $.

\begin{lemma}[Operational Correspondence, \cite{petersGlabbeek15}]
	An encoding $ \encoding : \procS \to \procT $ is operational corresponding \wrt a preorder $ {\relT} \subseteq \procT^2 $ that is a bisimulation iff $ \exists \IRel \logdot \left( \forall S \logdot \left( S, \Encoding{S} \right) \in \IRel \right) $\linebreak $ \wedge {\relT} = \Red{\IRel}{\procT} \wedge \left( \forall S, T \logdot \left( S, T \right) \in \IRel \text{ implies } \left( \Encoding{S}, T \right) \in {\relT} \right) \wedge \IRel $ is a preorder and a bisimulation.
\end{lemma}

\noindent
Here, the conditions $ \RelT = \Red{\IRel}{\procT} $ and $ \left( \forall S, T \logdot \left( S, T \right) \in \IRel \rightarrow \left( \Encoding{S}, T \right) \in \RelT \right) $ are technical side conditions for the proof of the only-if-part that ensure that $ \relT $ is a bisimulation.
We obtain the same result if we replace operational correspondence by strong operational correspondence and bisimulation by strong bisimulation.
Accordingly, operational correspondence ensures that source terms and their translations are reduction bisimilar.

To obtain a similar result for weak operational correspondence, \cite{petersGlabbeek15} had to introduce a new kind of simulation relation denoted as correspondence simulation.

\begin{lemma}[Weak Operational Correspondence, \cite{petersGlabbeek15}]
	$ \encoding $ is weakly operational corresponding \wrt a preorder $ {\relT} \subseteq \procT^2 $ that is a correspondence simulation iff $ \exists \IRel \logdot \left( \forall S \logdot \left( S, \Encoding{S} \right) \in \IRel \right) \wedge {\relT} = \Red{\IRel}{\procT} $\\ $ \wedge \; \left( \forall S, T \logdot \left( S, T \right) \in \IRel \text{ implies } \left( \Encoding{S}, T \right) \in {\relT} \right) \wedge \IRel $ is a preorder and a correspondence simulation.
\end{lemma}

\noindent
We omit the definition of correspondence simulation but point out that it is a simulation relation that is in between coupled similarity (Definition~\ref{def:coupledSimulation}) and bisimulation. Accordingly, weak operational correspondence ensures that source terms and their literal translations are coupled similar.

As stated in \cite{petersGlabbeek15}, the combination of the above results implies that the three semantic criteria of \cite{gorla10} ensure that any ``good'' encoding in this framework relates source and target terms by a coupled simulation that reflects divergence and respects success.
The approach presented in \cite{fuLu10, fu16} was not addressed in \cite{petersGlabbeek15} but it requires itself a simulation relation between source and target terms. Subbisimilarity is a variant of bisimulation. Accordingly, the requirements of \cite{fuLu10, fu16} are more restrictive than \cite{gorla10}.
The analysis of structural criteria is left for further research, because structural criteria need more assumptions on the considered languages.
As discussed in \cite{petersGlabbeek15}, the formal analysis of encodability criteria can also help to derive proof methods for the respective criteria.


\section{Conclusions}
\label{sec:conclusions}

As stated in the end of Section~\ref{sec:encodabilityCriteria}, there are basically three lines of further research:
\begin{inparaenum}[(1)]
	\item We need techniques to reason about the quality of encodability criteria to ensure that our encodings are indeed meaningful. Section~\ref{sec:evaluation} revises one way to do that, but raises itself some open questions as \eg how to deal with structural criteria.
	\item In order to compare results and build hierarchies we need a general notion of the quality of an encoding. Section~\ref{sec:frameworks} outlines three different frameworks for this purpose. All three approaches have certain basic ideas in common, \eg all three variants imply structural and semantic criteria. But the existence of three frameworks shows that there is still no general agreement and indeed these three frameworks differ not only \wrt the semantic equivalence they induce (see Section~\ref{sec:evaluation}) they also use quite different terminologies and basic definitions. With that they impose different proof techniques.
	\item Besides the analysis of encodability criteria, we are also lacking a study of proof techniques for particular criteria.
\end{inparaenum}


\bibliographystyle{eptcs}
\bibliography{Encodings}

\end{document}